\documentclass[iop,apj,twocolappendix,numberedappendix,appendixfloats]{emulateapj}
\pdfoutput=1 

\usepackage[breaklinks,colorlinks,citecolor=cyan, linkcolor=magenta]{hyperref} 
\usepackage{apjfonts} 
\usepackage{amsmath}

\usepackage{relsize}
\usepackage{amsbsy}
\usepackage{natbib} 
\usepackage{enumitem} 
\usepackage{mathtools}
\usepackage{lipsum}
\usepackage{xcolor}

\usepackage{graphicx}
\usepackage{wrapfig}

\shortauthors{Sevilla et al.}

\begin{document}
\received{}
\accepted{}


\title{Long-Term Lithium Abundance Signatures following Planetary Engulfment}



\author{Jason Sevilla\altaffilmark{1}, Aida Behmard\altaffilmark{2}, Jim Fuller\altaffilmark{1}}
\affil{$^{1}$Astronomy Department, California Institute of Technology, Pasadena, CA 91125, USA}
\affil{$^{2}$Division of Geological and Planetary Sciences, California Institute of Technology, Pasadena, CA 91125, USA}




\begin{abstract}
Planetary engulfment events can occur while host stars are on the main sequence. The addition of rocky planetary material during engulfment will lead to refractory abundance enhancements in the host star photosphere, but the level of enrichment and its duration will depend on mixing processes that occur within the stellar interior, such as convection, diffusion, and thermohaline mixing. We examine engulfment signatures by modeling the evolution of photospheric lithium abundances. Because lithium can be burned before or after the engulfment event, it produces unique signatures that vary with time and host star type. Using MESA stellar models, we quantify the strength and duration of these signatures following the engulfment of a 1, 10, or 100 $M_{\oplus}$ planetary companion with bulk Earth composition, for solar-metallicity host stars with masses ranging from 0.5$-$1.4 $M_{\odot}$. We find that lithium is quickly depleted via burning in low-mass host stars ($\lesssim 0.7 \, M_\odot$) on a time scale of a few hundred Myrs, but significant lithium enrichment signatures can last for Gyrs in G-type stars ($\sim \! 0.9 \, M_{\odot}$). For more massive stars (1.3$-$1.4 $M_{\odot}$), engulfment can enhance internal mixing and diffusion processes, potentially decreasing the surface lithium abundance. Our predicted signatures from exoplanet engulfment are consistent with observed lithium-rich solar-type stars and abundance enhancements in chemically inhomogeneous binary stars.

\end{abstract}



\section{Introduction} \label{sec:intro}
Refractory element abundances in stellar photospheres can be used as tracers for astrophysical events. Lithium is a particularly useful tracer because it is consumed in thermonuclear reactions at relatively low temperatures ($T \approx 3 \times 10^6 \, {\rm K})$ compared to other refractory species, and is thus depleted over stellar lifetimes. This implies that lithium abundances can shed light on recent events that altered stellar chemistry beyond birth compositions.

Photospheric lithium is depleted over time by convection and other stellar interior processes that can mix it down into the lithium burning region. Observations (see summary in \citealt{somers2016}) show that lithium abundances for such stars evolve and become depleted on timescales of millions to billions of years, and are thus affected by stellar interior processes acting throughout main sequence (MS) lifetimes. \citet{dumont2021} found that replicating such MS lithium depletion requires consideration of non-standard mixing processes. Thus, predictions of how long certain lithium abundance patterns will persist are strongly dependent on which mixing processes are involved, e.g., convection, thermohaline mixing, gravitational settling, element diffusion, mixing via shear instabilities, etc. 


Observed lithium abundance patterns may also be affected by planet formation and evolution processes, such as planetary engulfment. However, the evolution of lithium enrichment signatures resulting from engulfment events is not totally understood. \citet{sandquist2002} conjectured that lithium enrichments may indicate stellar pollution resulting from planetary engulfment. \citet{soaresfurtado2021} found significant lithium enrichment after modeling planetary engulfment events, but they did not consider effects from diffusion, overshoot mixing, and thermohaline mixing. On the other hand, \citet{theado2012} included thermohaline mixing, and found that engulfment could actually deplete surface lithium abundances below pre-engulfment levels.  As we shall see, these processes can alter surface abundances by orders of magnitude and deserve careful consideration.  

If planetary engulfment does affect photospheric lithium abundances, we might expect to see a difference in lithium observations of planet host stars compared to stars lacking known planets. \citet{israelian2009} observed lithium depletion in solar-type stars that harbor exoplanets compared to field stars. These lithium patterns may have resulted from prior engulfment of inner planets, or from sequestration of lithium within observed planets. Conversely, \citet{baumann2010} performed statistical tests to examine lithium abundances between two stellar populations, namely metal-rich solar analogues with and without observed exoplanets. They found no significant difference between the two samples, casting doubt on possible correlations between lithium abundances and planetary engulfment.



One particular type of mixing, thermohaline instability, is expected to be especially strong after engulfment events have occurred \citep{vauclair2004,garaud2011,bauer2018,bauer2019}. This is because thermohaline mixing is driven by an inverse mean molecular weight gradient, which is present after engulfment when heavy planetary material is deposited within outer layers of the engulfing star. Thermohaline mixing could attenuate lithium enrichment over time, or even cause lithium depletion below the primordial level.
This mixing process was the cause of the lithium depletion signatures observed by \citet{theado2012}. While these results are valuable, they cannot be extrapolated to a wide range of stellar types because \citet{theado2012} focused primarily on solar-type stars. 


To elucidate the connection between lithium enrichment and planetary engulfment while considering relevant mixing processes, we ran stellar models with the \texttt{MESA} stellar evolution code \citep{paxton2011,paxton2013,paxton2015,paxton2018,paxton2019}. In Section \ref{sec:model}, we discuss our \texttt{MESA} stellar models and implementation of non-standard mixing processes such as thermohaline instabilities. In Section \ref{sec:results}, we present analysis of the lithium abundances following planetary engulfment when different mixing processes are present, with varying amounts of accreted mass, and in comparison to models without engulfment. We summarize observable engulfment signature timescales in Section \ref{sec:timescale}, and discuss their implications for the evolution of planet engulfment signatures in Section \ref{sec:discussion}.  


\section{Stellar Models} \label{sec:model}


We computed our stellar models using the open-source 1D stellar evolution code \texttt{MESA} \citep{paxton2011,paxton2013,paxton2015,paxton2018,paxton2019}. This allows us to simulate mixing processes in the stars after planetary engulfment and to monitor surface lithium abundance over time. We ran non-rotating stellar models with zero-age main sequence (ZAMS) masses of 0.5--1.4 $M_{\odot}$, with solar metallicities of $Z$ = 0.017. Models that included accretion were run in three stages. In the first stage, we evolved the star up to its ZAMS phase. Planetary engulfment was simulated via accretion of bulk Earth composition material in the second stage (see Section \ref{sec:accrete}). 
In the third stage, we evolved the star up to the end of its MS lifetime. The surface lithium abundance in this final stage is the data of primary interest. We also ran models without accretion for comparison. In the first stage, convection is the only mixing process active.  In the second and third stages, each model utilized a selection of mixing settings that included convective overshoot mixing, thermohaline mixing, atomic diffusion, and a minimum \texttt{D\_mix} coefficient to model additional relevant (but poorly understood) mixing processes.



\subsection{Input Physics}\label{sec:mixing_desc}
Convective overshoot represents mixing that occurs near the convective boundaries of the star. We used an exponential scheme, which MESA takes from \citet{herwig2000}. The formula for the overshoot mixing coefficient is given in \citet{paxton2011}:

\begin{equation}
D_{\rm ov} = D_{\rm conv, 0}\hspace{0.7mm} \exp  \left ( -\frac{2z}{f \lambda_{P, 0}}\right) , \hspace{1mm}
\end{equation}

\noindent where $D_{\rm conv, 0}$ the diffusion coefficent taken from a point determined by the user, $\lambda_{P, 0}$ is the pressure scale height at this point, $z$ is the distance away from this point, and $f$ determines the characteristic size of the overshooting region in terms of $\lambda_{P, 0}$. The settings in the MESA inlist allow us to input $f$ as well as $f_0$, which is how many scale heights into the convective zone the aforementioned point will be placed to calculate $D_{\rm conv, 0}$. We chose $f=0.02$ and $f_0=0.005$ in all of our runs.

Thermohaline mixing is a double diffusive instability that occurs in the presence of an inverse mean molecular weight gradient and a stabilizing entropy gradient, i.e., material of higher mean molecular weight and entropy lying above material of lower mean molecular weight and entropy. The material with high mean molecular weight sinks downwards in long strands, and material with lower mean molecular weight rises up.
This type of mixing is especially relevant after planetary engulfment, where heavy planetary material is deposited near the stellar surface. The diffusive mixing from thermohaline has been calibrated from numerical simulations (e.g., \citealt{denissenkov2010,traxler2011,harrington2019}).

We utilize \texttt{MESA}'s thermohaline mixing prescription based on \citet{brown2013} in our \texttt{MESA} models. \citet{brown2013} provides a more accurate prescription of thermohaline mixing compared to previous implementation (e.g., \citealt{kippenhahn1980}) via an improved fingering convection model that is supported by 3D numerical simulations (\citealt{zemskova2014}).  The thermohaline instability is only active when the composition gradient $\nabla_\mu = d \ln \mu/d \ln P$ is sufficiently negative, and the effective thermohaline diffusivity $D_{\rm th}$ approaches zero when $\nabla_\mu$ approaches the stability boundary (see discussion in \citealt{bauer2019}). 

The prescription for how \texttt{MESA} handles elemental diffusion is provided in \citet{paxton2018}, using Burgers' diffusion equations. All isotopes are assigned to a diffusion class that covers a range of atomic masses. Each isotope in a class is treated identically to that class’s representative isotope. These diffusion classes are summarized in Table \ref{tab:table1}.

\begin{deluxetable}{cc}
\setlength{\tabcolsep}{1em}
\tablewidth{0.27\textwidth}
\tabletypesize{\footnotesize}
\tablewidth{0pt}
\tablecaption{List of representative isotopes and atomic mass ranges for each diffusion class. \label{tab:table1}}
\tablecolumns{2}
\tablehead{
\colhead{Representative Isotope} &
\colhead{Atomic Mass Range}
}
\startdata
$^{1}$H & $1-2$ \\
$^{4}$He  & $3-4$ \\
$^{7}$Li & $5-7$ \\
$^{12}$C & $8-12$ \\
$^{16}$O & $13-16$ \\
$^{24}$Mg & $17-24$ \\
$^{28}$Si & $25-28$ \\
$^{56}$Fe  & $>28$ \\
       \vspace{-2.5mm}
       
\end{deluxetable}

\begin{figure}
\centering
    \hspace*{-0.5cm}
    \includegraphics[width=0.54\textwidth]{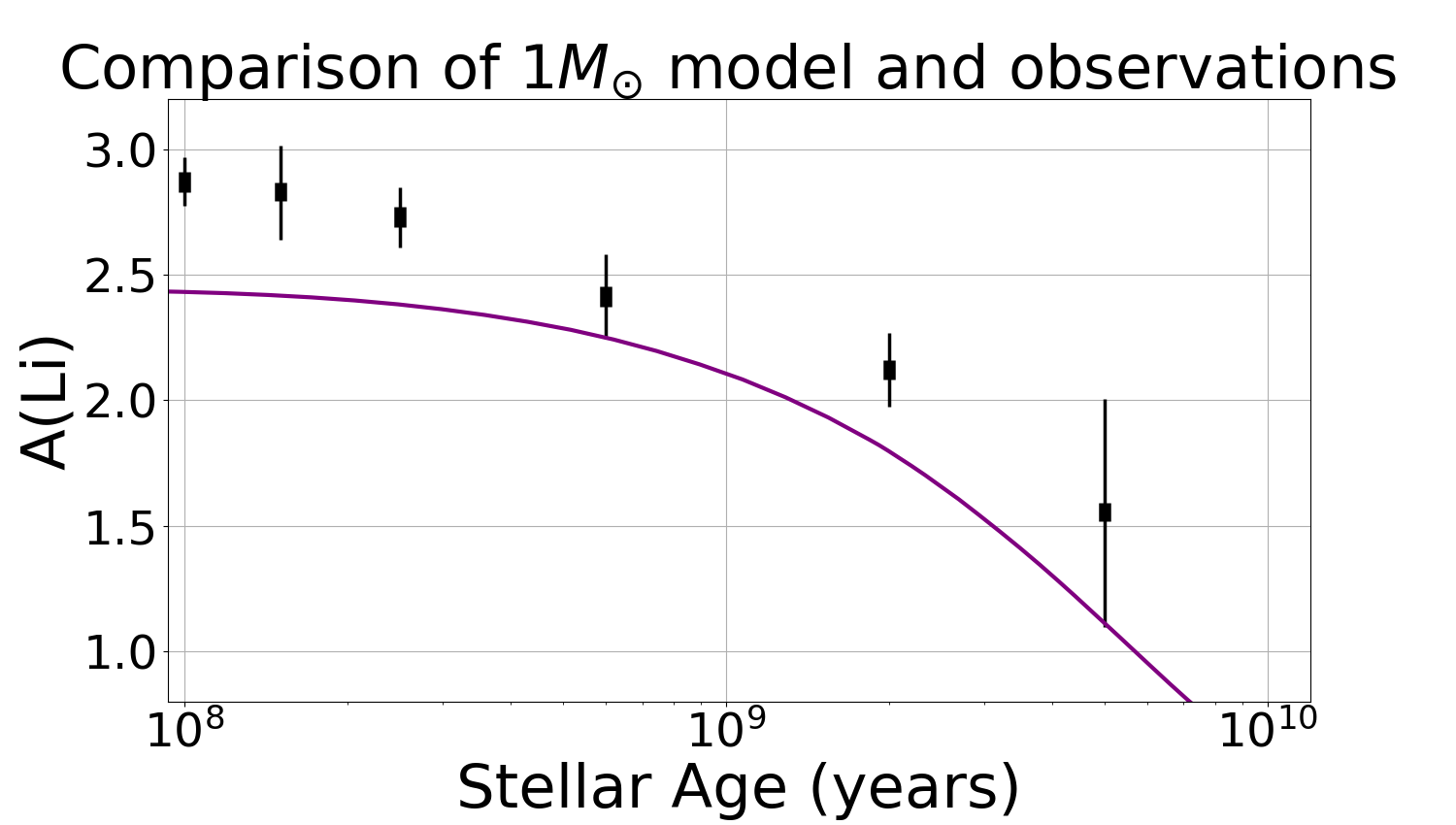}
\caption{\footnotesize{Comparison of the \texttt{MESA} lithium abundance for a 1 $M_{\odot}$ (purple line) without engulfment and observed lithium abundances of stars in nearby clusters
from \citet{somers2016} (black points).}} 
\label{fig:figure1}
\end{figure}

There are other mixing effects at play that we did not consider, such as rotationally induced mixing. To account for these other mixing processes, we compared the MS evolution of the surface lithium abundance for a 1 $M_{\odot}$ star with the observed lithium abundances of solar-like stars provided in \citet{somers2016}, and added a minimum diffusivity coefficient, \texttt{D\_mix}, to reproduce the observations. Specifically, the value of this coefficient was adjusted so that the level of lithium depletion in a 1 $M_{\odot}$ model fits the observed data from \citet{sestito2005}, as seen in Figure \ref{fig:figure1}.
Our model has slightly less lithium than typical stars, and we attribute this offset to poorly understood mixing processes in the protostar that set the lithium abundance at ZAMS. However, we are only interested in the level of lithium depletion relative to an initial value. So, we adjusted \texttt{min\_D\_mix} so our model would have the appropriate level of surface lithium depletion relative to initial lithium abundance.  This yielded a \texttt{min\_D\_mix} value of 700, which we applied to all of our models.

\begin{figure*}[t]
\centering
     \centering
        \vspace{0mm}
        \includegraphics[width=0.99\textwidth]{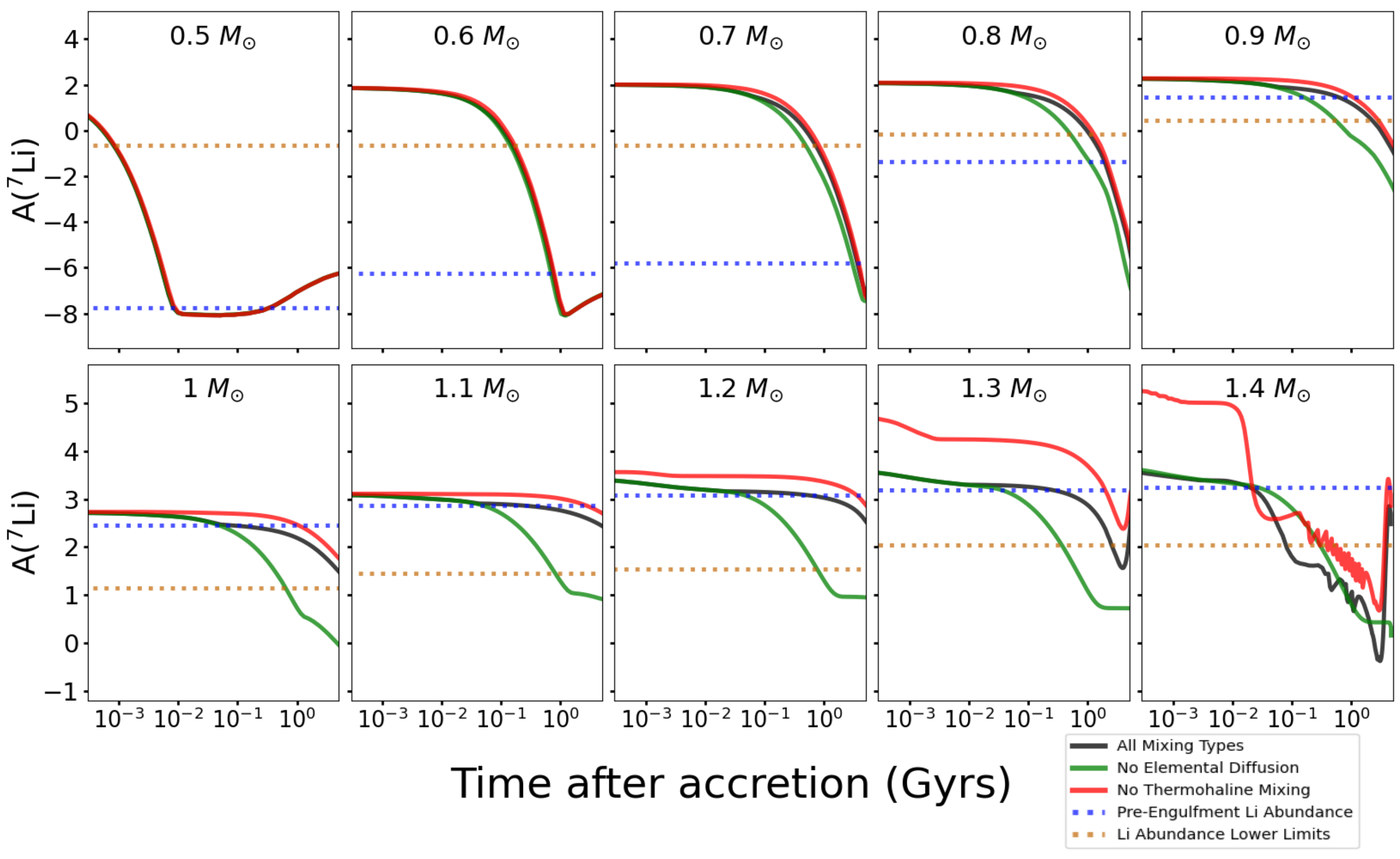} 
\caption{\footnotesize{The lithium abundances in dex following a 10 M$_{\oplus}$ engulfment} as a function of time for the three sets of models over the stellar mass range of 0.5$-$1.4 $M_{\odot}$. The black line represents the model with all the mixing processes described in section \ref{sec:mixing_desc}, the red line excludes thermohaline mixing, and the green line excludes elemental diffusion. The dashed blue line is the lithium abundance before the engulfment event, and the dashed yellow line is the approximate lithium detection threshold.}
\label{fig:figure2}
\end{figure*}

\subsection{Model Setup and Accretion}
\label{sec:accrete}

The star was initialized with a metallicity of $0.02$ as that is the only possible value for a default \texttt{MESA} installation. Then we gradually adjusted the metallicity to $0.017$, to be closer to solar metallicity. This set the star's initial (pre-ZAMS) lithium abundance to be at A($^7$Li)$\sim$3.3 for all models. 
During pre-MS evolution, we disabled all mixing processes other than convection, as we found the use of overshoot depleted the lithium abundance by roughly two orders of magnitude compared to observed for solar-like stars in Figure \ref{fig:figure1}. We discuss this effect further in Section \ref{sec:limitations}.
The lithium abundance at the time of ZAMS is displayed as the leftmost intercept of the pink lines in Figure \ref{fig:figure5}. ZAMS is defined as when the proportion of total luminosity from nuclear burning reaches 0.99 or greater.

At ZAMS, we simulated planetary engulfment as accretion of 1, 10, or 100 $M_{\oplus}$ with bulk Earth composition. In reality, planetary engulfment is a rapid process that should be completed within timescales of years. However, rapid accretion prevented the \texttt{MESA} models from successfully iterating.  So, we set accretion to occur at $10^{-9} M_{\odot}$ per year, or $3\times10^{-4} M_{\oplus}$ per year. This should not affect our results as the accretion period is still relatively short compared to the timescale of the lithium depletion we observed in our models (Table \ref{tab:table3}).

To facilitate accretion, we added extra mixing to the outermost $10^{-3}$ percent of the stellar mass.
This extra mixing helps spread the accreted material throughout the stellar outer layers, preventing it from accumulating at the outermost grid point.
This allows MESA to run faster. To prevent the model from overshooting the target accreted mass by taking too large of a time step, we set the maximum time step to be $10^9$ seconds per total accreted Earth mass.

To accrete material of bulk Earth composition, we included the ten most prevalent elements plus $^7$Li as represented by their most common isotope. There are two stable isotopes of lithium, $^{6}$Li and $^{7}$Li. We chose to focus on $^{7}$Li for two reasons: $^{7}$Li is the more abundant isotope, and it has a relatively higher burning temperature, so $^6$Li is already mostly depleted before stars reach the MS \citep{sandquist2002}. The observed abundance is thus dominated by $^7$Li in most cases.

Each isotopic mass abundance was taken from \citet{mcdonough1995}. Because we excluded some elements, the mass fractions
sum to less than 100\%. We accounted for this by adding extra material of the most common bulk Earth element, iron. The resulting elemental abundances in the accreted planet are given in Table \ref{tab:table2}.  The mean molecular weight of these updated abundances are within 1\% of the mean molecular weight of the original abundances.
We set the accretion event to occur at ZAMS, around the time of planet formation, because we expect that many engulfments will likely occur during this early period when planetary systems are relatively unstable. 
After accretion was complete, we evolved the star to the end of its MS lifetime. We used default \texttt{MESA} definition for the end of the MS, which is when the core hydrogen mass fraction drops below $10^{-10}$. 

\begin{deluxetable}{cc}
\setlength{\tabcolsep}{1em}
\tablewidth{0.27\textwidth}
\tabletypesize{\footnotesize}
\tablewidth{0pt}
\tablecaption{Mass fractions of each isotope in the engulfed planet. \label{tab:table2}}
\tablecolumns{2}
\tablehead{
\colhead{Element} &
\colhead{Mass Fraction (\%)}
}
\startdata
$^{56}$Fe & 32.40289 \\
$^{16}$O  & 29.7 \\
$^{28}$Si & 16.1 \\
$^{24}$Mg & 15.4 \\
$^{58}$Ni & 1.822 \\
$^{40}$Ca & 1.71 \\
$^{27}$Al & 1.59 \\
$^{32}$S  & 0.635 \\
$^{52}$Cr & 0.47 \\
$^{23}$Na & 0.18 \\
$^7$Li  & $1.1 \times 10^{-4}$  
\tablecomments{Bulk Earth composition taken from \citet{mcdonough1995}, and modified according to Section \ref{sec:accrete} to account for omitted elements.}
\end{deluxetable}

\section{Results}
\label{sec:results}

\subsection{Comparison of Differing Mixing Models} \label{sec:mixing_processes}

We examined the effects of overshoot mixing, thermohaline mixing, and elemental diffusion, with the goal of understanding the effects each mixing process has on the evolution of photospheric $^7$Li abundance as a function of stellar mass. 
Each model accreted 10 $M_{\oplus}$ of material. There were three sets of models, each including different mixing processes. The first had all of the mixing types in the previous section. The second model excluded thermohaline mixing, while the last excluded elemental diffusion.

Figure \ref{fig:figure2} shows the plots of MS $^7$Li surface abundance, grouped by initial stellar mass.
There are three regimes, based on the star’s mass. In the low-mass regime, from 0.5 to 0.8 $M_{\odot}$, the convective zone extends deep enough to reach near the lithium burning region. These models exhibit similar patterns of rapid lithium depletion due to efficient burning. Neither thermohaline mixing nor diffusion significantly change the effects of convective/overshoot mixing that drive lithium from the surface down to its burning region.

In the solar-like mass regime, from 0.8 to 1.2 $M_{\odot}$, we see that surface $^7$Li abundance depends more strongly on mixing processes. 
The models without thermohaline mixing have the least lithium depletion, while the models without elemental diffusion have the most lithium depletion. This indicates that thermohaline mixing is the most effective mixing process at depleting surface lithium.
Surprisingly, the models with diffusion/gravitational settling exhibit less lithium depletion than the models without diffusion/settling.
This upsets our expectation that diffusion/settling would contribute to lithium sinking into the radiative zone, and therefore more lithium depletion at the surface.

\begin{figure}
\centering
\hspace*{-0.4cm}
    \includegraphics[width=0.41\textwidth]{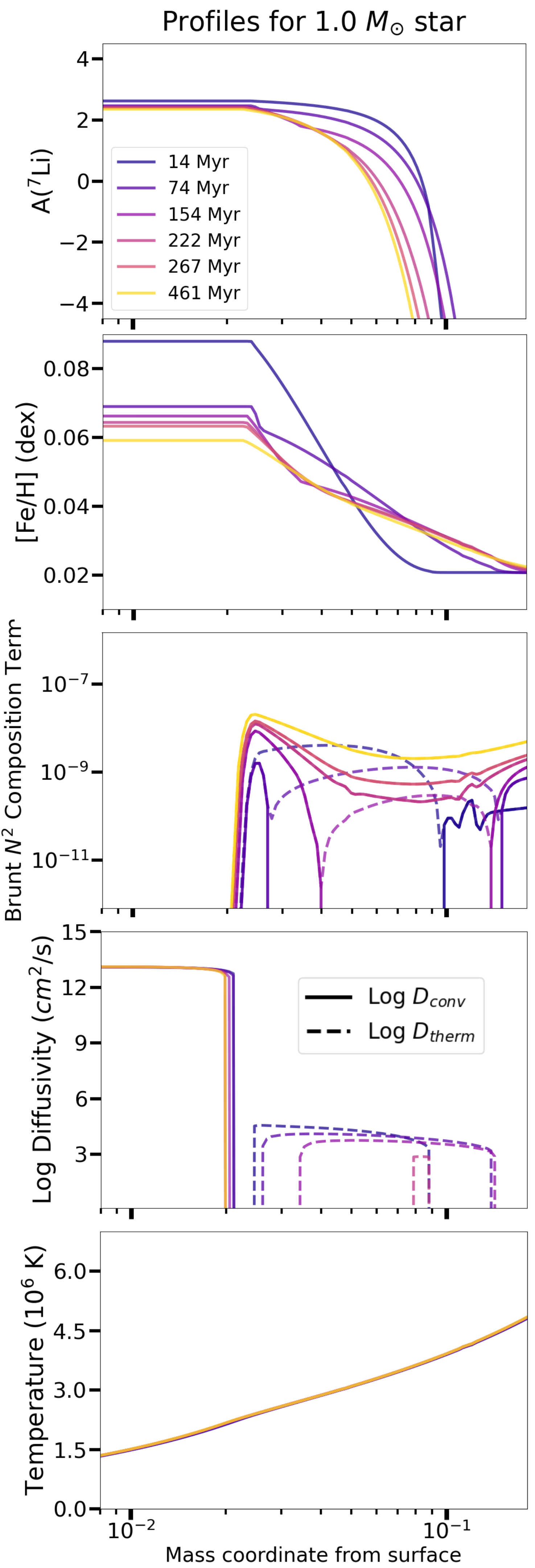}
    \vspace*{0in}
\caption{\footnotesize{As a function of mass below the surface, profiles of the internal $^7$Li abundance (top), iron abundance (second panel), the Brunt $N^2$ composition term (third panel), convective diffusivity $D_{\rm conv}$ and thermohaline diffusivity $D_{\rm therm}$ (fourth panel), and temperature (bottom) in a 1.0 $M_{\odot}$ star with all mixing processes enabled, following a 10 $M_{\oplus}$ engulfment. Lines are labeled by time after the engulfment event. A negative $N^2$ composition term is shown as dotted lines.
The inverse mean molecular weight gradient and thermohaline mixing disappear a couple hundred million years after engulfment.}}
\label{fig:figure3}
\end{figure}

The reason that diffusion/settling inhibits lithium depletion is the reliance of thermohaline mixing on an inverse mean molecular weight gradient. Gravitational settling allows heavier elements to settle below the convective zone. While this does cause some of the accreted elements (e.g., iron) to sink deeper into the star, the primary effect is to allow the star's helium to sink slightly deeper into the star. This creates a stabilizing mean molecular weight gradient, shutting off thermohaline mixing, as demonstrated in Figure \ref{fig:figure3}. This effect is also observed by \citet{theado2012}. That study observed the creation of a stable mean molecular weight gradient as a result of helium settling, which counteracted thermohaline mixing.

We verified this effect by tracking the compositional component of $N^2$ (the Brunt-V\"ais\"al\"a freqency squared). The $N^2$ composition term is
positive for a stable mean molecular weight gradient and negative when thermohaline mixing occurs.
Initially, the $N^2$ composition term is negative just under the convective zone, showing that there is an inverse mean molecular weight gradient that allows for thermohaline mixing.
On timescales of $\gtrsim$100 Myr after accretion, the $N^2$ composition term shifts to be positive under the convective zone due to gravitational settling, shutting off thermohaline mixing.
Therefore, thermohaline mixing is only active in the short timespan just after engulfment when gravitational settling has not yet formed this stabilizing gradient. The plots in Figure \ref{fig:figure2} display this effect. For example, for our 10 $M_{\oplus}$ accretions, the lithium depletion for the all mixing processes model (black line) is greater relative to a model without thermohaline mixing (red line) for the first $\sim \! 10^8$ years before a stable mean molecular weight gradient has formed via helium settling. Afterwards, these models mirror the path of the models without thermohaline mixing, as lithium is depleted more gradually by ongoing gravitational settling.



\begin{figure}
\centering
\hspace*{-0.5cm}
    \includegraphics[width=0.41\textwidth]{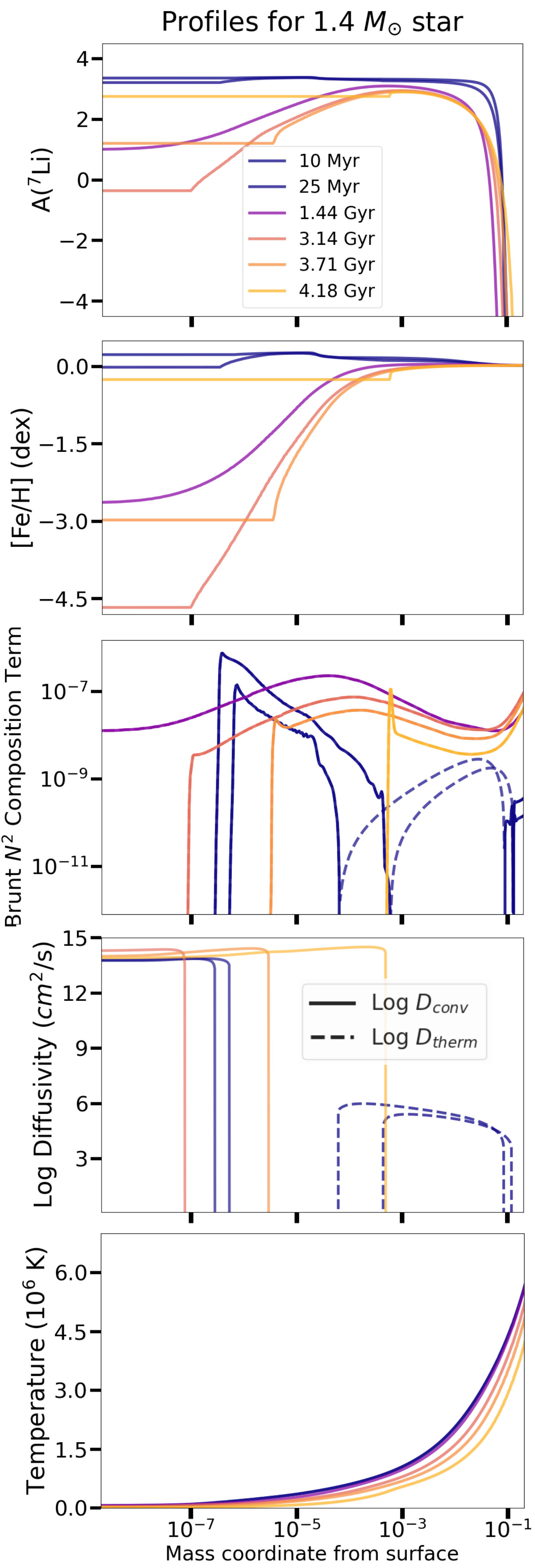}
    \vspace*{0in}
\caption{\footnotesize{Same as Figure \ref{fig:figure3}, but for a 1.4 $M_{\odot}$ star.
}}
\label{fig:figure4}
\end{figure}

\begin{figure*}
\centering
     \centering
        \vspace{0mm}
        \includegraphics[width=0.98\textwidth]{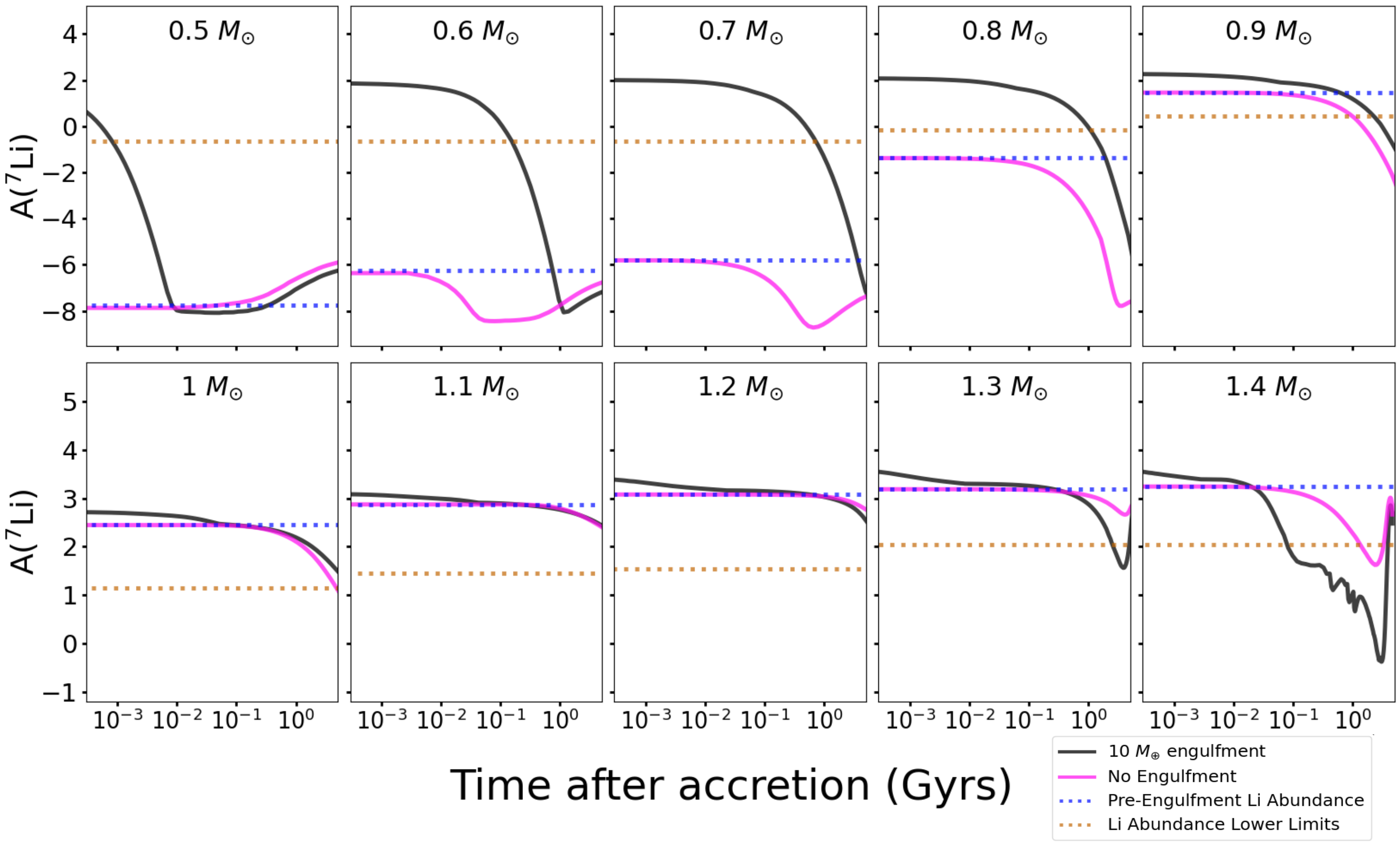}
\caption{\footnotesize{Surface lithium abundances as a function of time for models with (black line) and without (magenta line) a $10 \, M_\Earth$ planetary engulfment. Both models include all mixing processes. The dashed blue line represents the pre-engulfment lithium abundance, and the dotted yellow line indicates the approximate lower limit to the measurable lithium abundance.}}
\label{fig:figure5}
\end{figure*}

In the higher mass range, including the 1.3 and 1.4 $M_{\odot}$ stars, we see a new pattern. The model without diffusion behaves similarly as in the lower mass models, exhibiting a sustained decrease in surface lithium abundance that bottoms out as time goes on. But in the models that include elemental diffusion, the lithium abundance increases again after a few Gyr. This occurs because the surface convective zones are extremely thin in these more massive stars, but they deepen with time.

\begin{figure*}
\centering
    \includegraphics[width=0.98\textwidth]{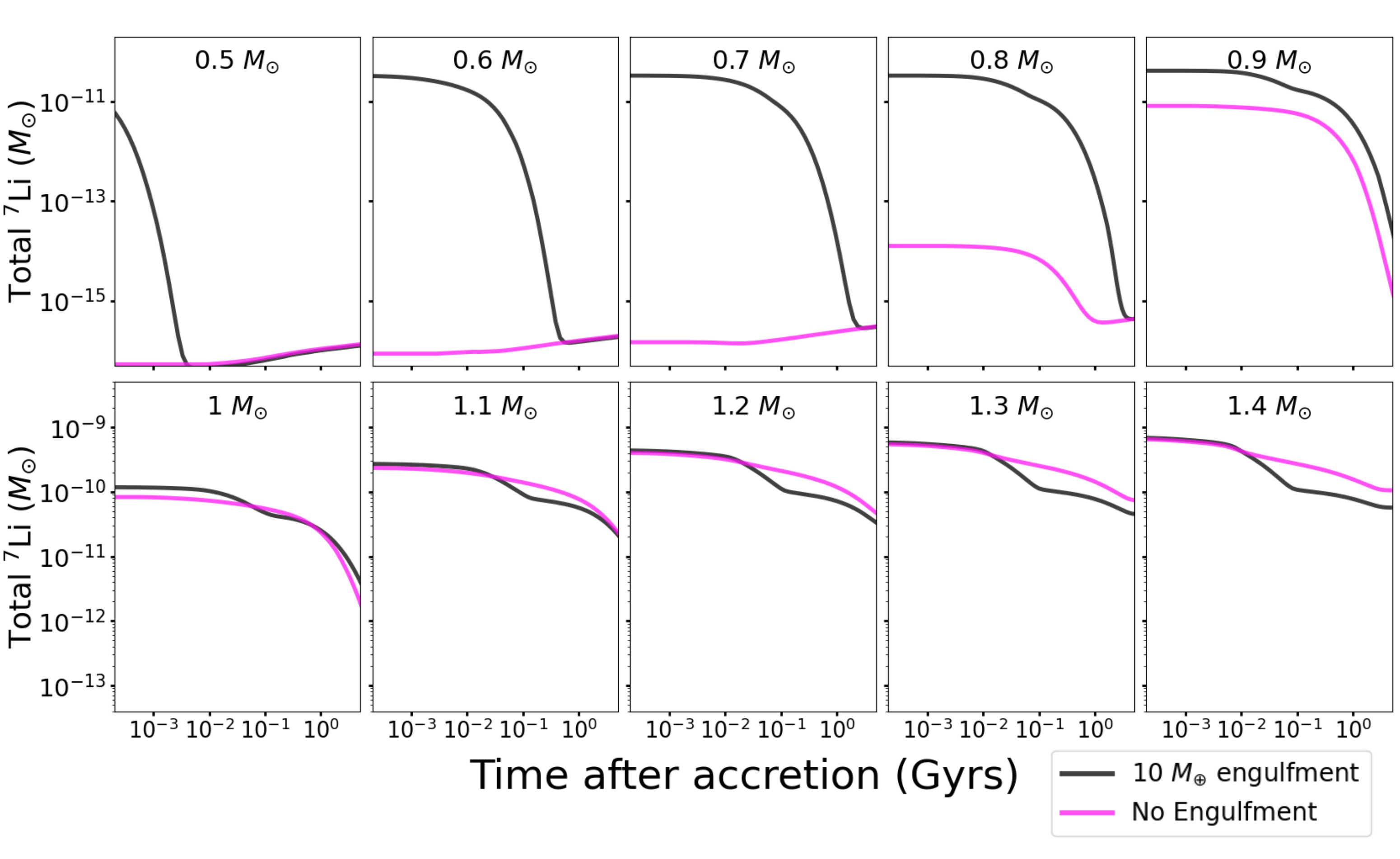}
    \vspace*{0.05in}
\caption{\footnotesize{Total mass of $^7$Li within the star for the 10 $M_\Earth$ engulfment models (black) and non-engulfment models (magenta).  Note the different y-axis ranges for the top and bottom rows.}}
\label{fig:figure6}
\end{figure*}

Initially, lithium and other metals settle underneath the star's surface convective zone, as
illustrated in Figure \ref{fig:figure4}. As in the case of solar-like stars, gravitational settling quickly establishes a stable molecular weight gradient under the convective zone, shutting off thermohaline mixing
and preventing lithium from being mixed downwards.
Much of the $^7$Li remains just underneath the convective zone, at a depth of $\sim \! 10^{-3} \, M_\odot$.
Towards the end of the MS, $\sim$4 Gyr after accretion, the convective zone deepens as the star approaches the end of the MS. Lithium and the other heavy elements are then mixed back up towards the surface. In models including only thermohaline mixing, lithium and other metals are mixed deeper into the star, and are not dredged up by the surface convection zone until after the MS.  

\subsection{Engulfment vs. No Engulfment}

We now compare the models with and without a planetary engulfment event, shown in Figure \ref{fig:figure5}. These models are for a 10 $M_{\oplus}$ accretion, but we will show that the patterns are similar for other planet masses in Section \ref{sec:diffmasses}.  
From 0.5 to 1.2 $M_{\odot}$,
the lowest mass stars ($M \lesssim 0.7 \, M_\odot)$ without planetary engulfment have the lowest lithium abundances because of prior depletion of lithium due to burning.

In models with engulfment, the surface lithium abundance is raised to $A(^7{\rm Li}) \sim 2$, several orders of magnitude larger than the low-mass non-engulfment models. 
However, the fast depletion in these mass stars indicates that lithium is depleted on the order of $10^7$ to $10^8$ years.

At the higher end of this mass range ($0.8-1.2 \, M_\odot$), lithium depletion is much slower.
 Lithium enhancements of a factor of $\sim \! 10^3$ can be sustained for Gyrs in $0.8 \, M_\odot$ stars, though the lithium abundance falls below detectable limits after $\sim$1 Gyr. With increasing stellar mass, the magnitude of lithium enhancement decreases, as the initial lithium abundance prior to engulfment is much higher in higher mass stars.  For example, the magnitude of lithium enrichment is typically less than a factor of 2 in the $1.0-1.2 \, M_\odot$ models. The absolute lithium abundance is highest in the $M \approx 1.2 \, M_\odot$ models, which sustain $A(^7{\rm Li} \sim 3)$ for more than a Gyr.  This is because the convective zone is smaller for more massive stars, and is thus further away from the lithium burning region, which reduces  the amount of lithium burning due to convection in the pre-MS.


This pattern changes for 1.3 and 1.4 $M_{\odot}$ stars, where models without planetary engulfment have more lithium than the models with engulfment.  
This occurs because the higher concentration of heavy elements from engulfment allows these elements to be mixed downward more efficiently through thermohaline mixing, followed by gravitational settling. The result is that relatively soon after the engulfment event ($\sim$30 Myr for the $1.4\, M_\odot$ model), the surface lithium abundance of an accreting model drops below its non-accreting counterpart. It remains lower throughout the MS, causing stars that underwent planetary engulfment to exhibit significantly lower surface lithium compared to stars that did not. 

In the models with engulfment, there is a phase of faster lithium depletion in the first 1$-$100 Myr. 
This phase is mentioned in Section \ref{sec:mixing_processes} when thermohaline mixing is active for a short time after engulfment due to the presence of heavy metals near the surface. This also explains the absence of this decrease in the non-engulfment models. Although the non-engulfment models include the same mixing processes, the lack of accretion means that there is no inverse mean molecular weight gradient to drive thermohaline mixing. The length of this phase is longest in the lower mass stars, as a larger convective zone increases the diffusion time scale at the base of the convective envelope, delaying the formation of a stabilizing mean molecular weight gradient via gravitational settling. Note that the length of this phase depends on the mass of the engulfed planet, which will be discussed in \ref{sec:diffmasses}.


    
    

\begin{figure*}
\centering
    \includegraphics[width=0.92\textwidth]{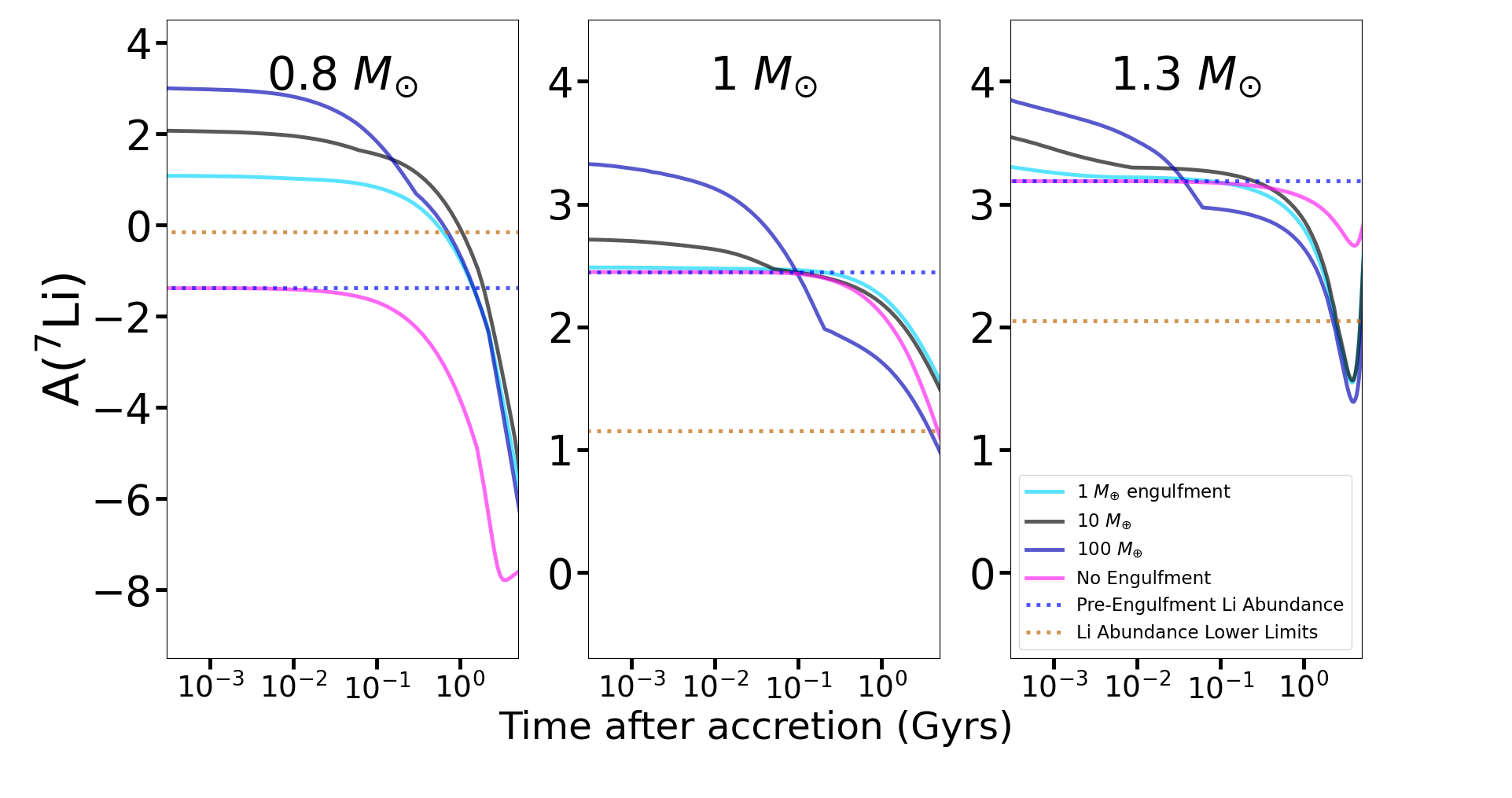}
\caption{\footnotesize{Surface lithium abundances following engulfment of different planetary masses for three stellar models. The light blue line represents an engulfment of 1 $M_{\oplus}$, the black line for 10 $M_{\oplus}$, and dark blue for 100 $M_{\oplus}$.  The magenta line represents a non-engulfment model, the dashed blue line shows the pre-engulfment lithium abundance, and the dashed yellow line is the approximate detection threshold. All models shown have all mixing types enabled.
}}
\label{fig:figure7}
\end{figure*}

\subsection{Total Lithium Mass}

We also examined the total mass of $^7$Li within the star to determine whether surface depletion occurs because of lithium burning, or just downward lithium mixing (Figure \ref{fig:figure6}).
As expected, low-mass stars strongly deplete the total $^7$Li mass
due to convective mixing to the lithium burning regions. In these stars, the trend of total lithium mass follows the trend of surface lithium depletion. Despite this burning, in stars with $M \! \lesssim \! 1.0 \, M_\odot$, the models with engulfment always retain more lithium than models without engulfment.

In higher mass stars, the total lithium mass decreases much less than the surface lithium abundance. In the 1.3 and 1.4 $M_{\odot}$ stars, the total lithium depletion  is about a factor of 3, which is less than the factor of $\sim$100 depletion in surface lithium depletion. This indicates that gravitational settling (which does not extend all they way to the burning region) is the primary driver behind the sharp surface lithium depletion in these stars, not lithium burning.
However, note that planetary accretion does stimulate enough thermohaline mixing to decrease the total mass of lithium below the pre-engulfment level for stars with $M \geq 1.2 \, M_\odot$ by $\sim$30 Myr after the engulfment.

\subsection{Accretion of Different Masses}
\label{sec:diffmasses}

We also examined models that engulfed planets of 1 $M_{\oplus}$ and 100 $M_{\oplus}$. We compared these to our previous models of 10 $M_{\oplus}$ engulfment and no engulfment. Figure \ref{fig:figure6} shows these surface lithium abundances for stars of $0.8$, $1.0$, and $1.3 \, M_\odot$. Like the 10 $M_{\oplus}$ engulfment models, these models had all mixing processes enabled.

As we would expect, the models that engulf more mass start with a higher lithium abundance. Higher stellar masses have smaller convective zones, meaning they begin with the highest surface lithium abundances. At the same time, the higher engulfed planetary mass means that there is a stronger inverse mean molecular weight gradient. This leads to stronger thermohaline mixing right after engulfment, as evidenced by the steeper initial drops in lithium abundance in the models with larger engulfed planetary masses. These periods of thermohaline mixing also last longer as the star engulfs more mass: in a $1 \, M_\odot$ star, the thermohaline mixing lasts for approximately 5 Myr years for the 1 $M_{\oplus}$ engulfment, 50 Myr years for 10 $M_{\oplus}$, and 200 Myr years for 100 $M_{\oplus}$. 

In the 100 $M_{\oplus}$ engulfment models, the extended period of strong thermohaline mixing is enough to wipe out the extra surface lithium abundance above the lower mass engulfment models. In some cases, the lithium abundance of the 100 $M_{\oplus}$ engulfment falls below the lithium abundance of the 1 and 10 $M_{\oplus}$ engulfment models. The resulting effect is that the lithium abundances of the various engulfment models tend to converge no matter how much mass is accreted. This indicates that there is a limit to the amount of achievable lithium enrichment via planetary engulfment more than $\sim \! 10^8$ yr after the engulfment event, and that inferring the mass of the engulfed planet is very difficult.

 
\subsection{Different Accretion Time}
\label{sec:1gyr}

\begin{figure*}
\centering
    \includegraphics[width=0.98\textwidth]{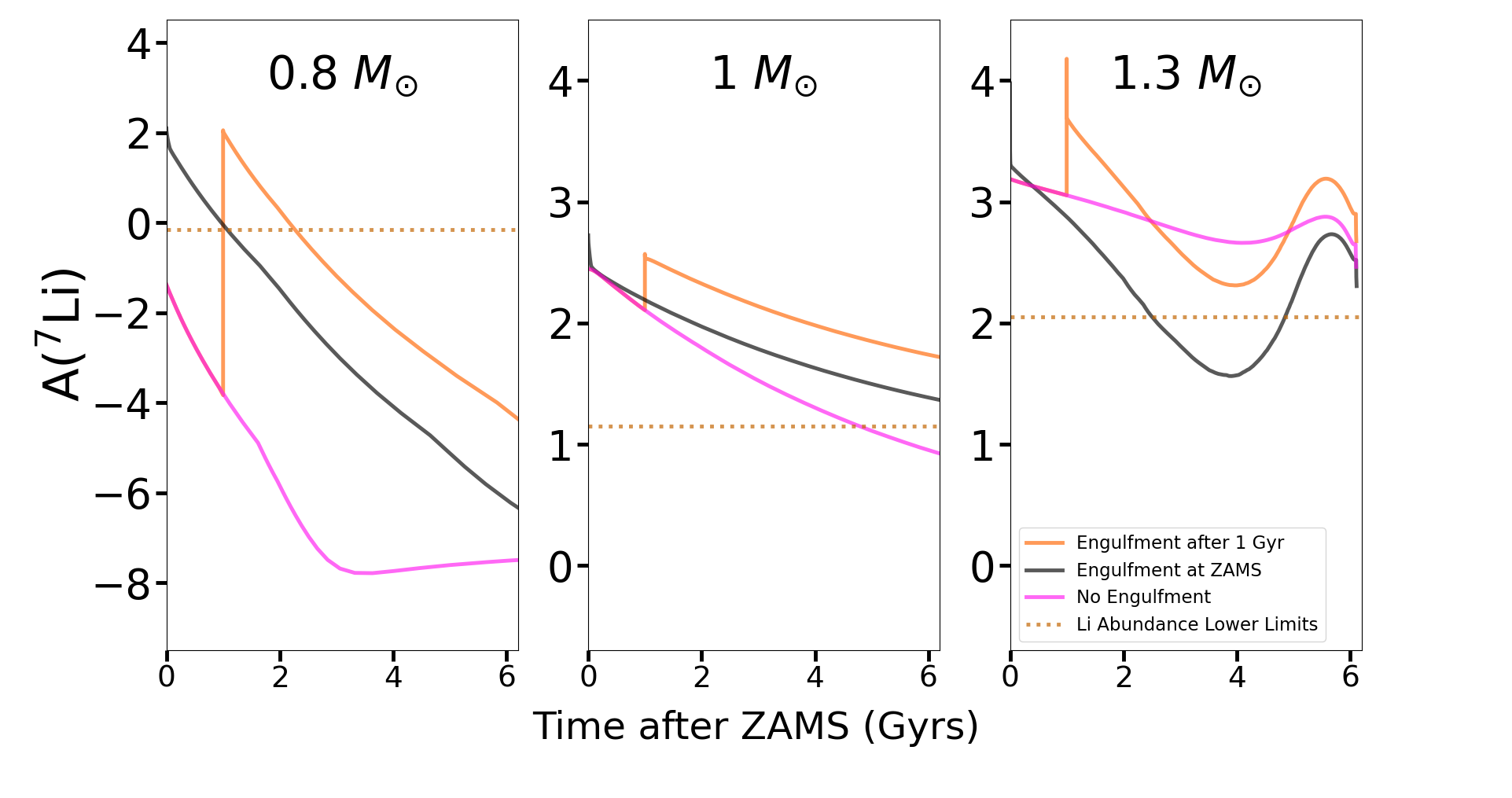}
\caption{\footnotesize{Surface lithium abundance for a 10 $M_\Earth$ engulfment at ZAMS (black), a 10 $M_\Earth$ engulfment 1 Gyr after ZAMS (orange) and the non-engulfment model (magenta).  The orange dashed line shows the detection threshold for lithium abundance.  All models include all mixing processes in Section \ref{sec:mixing_processes}. The x-axis is in linear time scale to emphasize long-term behavior.}}
\label{fig:figure8}
\end{figure*}

Due to the interplay between thermohaline mixing and gravitational settling, the timing of the engulfment event can affect the surface lithium enhancement. Results for engulfment events occurring 1 Gyr after ZAMS are shown in Figure \ref{fig:figure8} alongside the ZAMS engulfment model and the control model without engulfment.
For low-mass stars ($M \! \approx \! 0.8 \, M_\odot$), the evolution of $^7$Li following the engulfment 1 Gyr after ZAMS closely tracks the ZAMS engulfment model, and is only shifted 1 Gyr later. In both cases, the lithium enhancement is observable for roughly 1.5 Gyr before the lithium abundance drops below detectable limits.

The behavior is similar for solar-mass stars, but with an important difference. The post-engulfment lithium depletion caused by thermohaline mixing is smaller when the engulfment occurs 1 Gyr after ZAMS. This occurs because thermohaline mixing is suppressed by the stabilizing composition gradient that has formed due to helium settling throughout the first Gyr of the star's life.  Additionally, lithium is partially depleted during the first Gyr of evolution, so the planet engulfment increases the lithium abundance by a larger factor after 1 Gyr than it does at ZAMS. Consequently, the 1 Gyr engulfment model has consistently higher surface lithium abundance than the ZAMS engulfment model at comparable times after the engulfment event.  

For models with a stellar mass $\approx \! 1.3 \, M_{\odot}$, the model with engulfment 1 Gyr after ZAMS follows the same trends as the ZAMS engulfment model, but with significantly higher surface lithium abundance, again due to suppressed thermohaline mixing due to helium settling. Depending on the time after engulfment, the surface lithium abundance could be slightly larger or slightly smaller than a star without planetary engulfment.

\subsection{Comparison with Iron Abundance}

\begin{figure*}
\centering
    \includegraphics[width=0.98\textwidth]{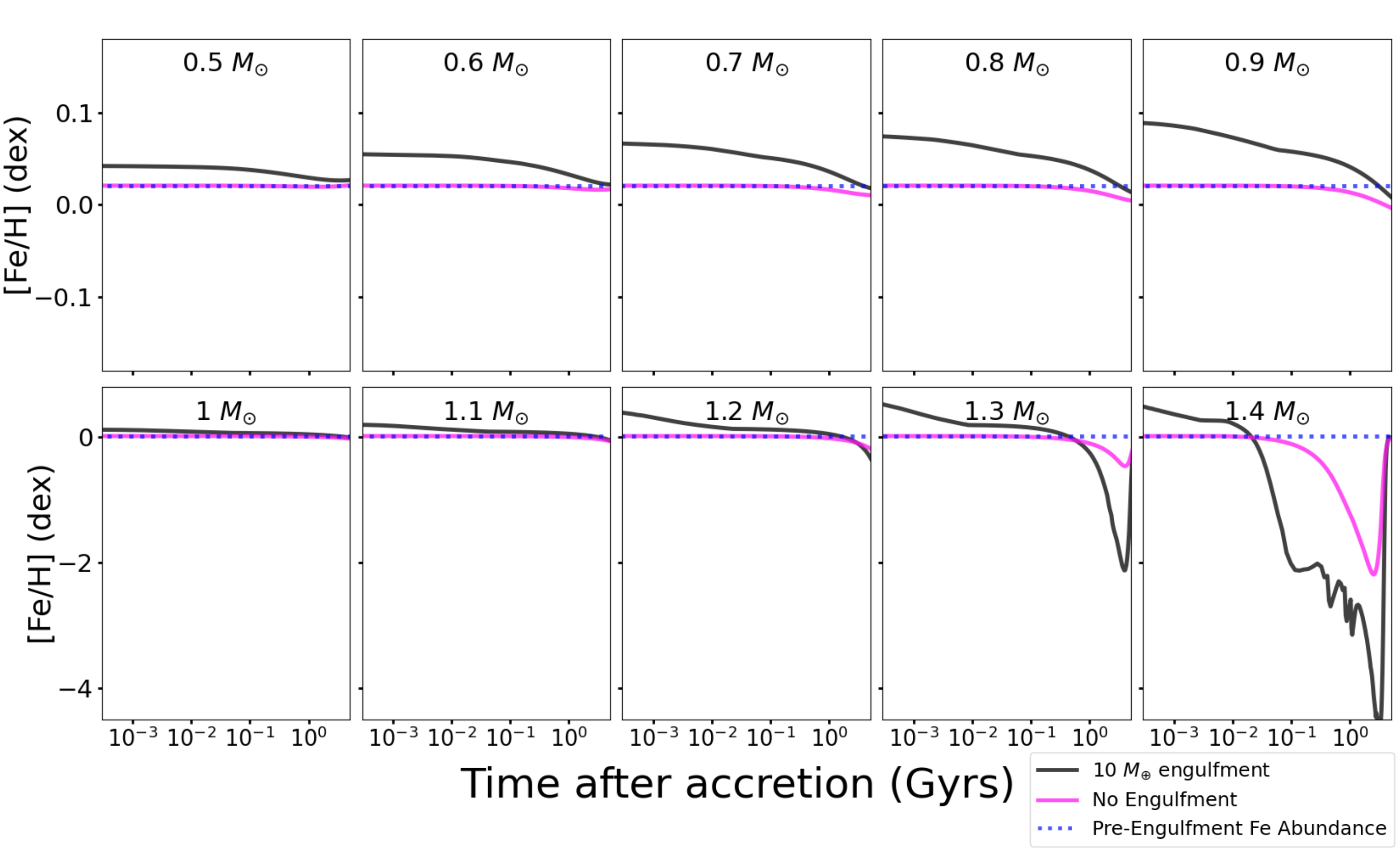}
\caption{\footnotesize{Surface iron abundance for the 10 $M_\Earth$ engulfment models (black) and the non-engulfment model (magenta).  The blue dashed line shows the iron abundance prior to engulfment.  Note the different y-axis range in the top and bottom rows.}}
\label{fig:figure9}
\end{figure*}

We can briefly compare these patterns of lithium abundance with those of iron abundance to investigate why lithium is unique amongst other chemical tracers. We find that the abundances of all metals not including lithium (e.g., iron, silicon, magnesium, etc.) behave in a similar fashion. Figure \ref{fig:figure9} shows the surface iron abundance of our models. In models above 1.1 $M_{\odot}$, surface iron abundance follows a similar trend to surface lithium abundance due to thermohaline mixing and gravitational settling.
The increase in surface iron abundance in $M \! \lesssim \! 1\, M_\odot$ stars due to planetary accretion is less than the increase in lithium abundance because the star already contains a great deal of iron, which is not destroyed during the pre-main sequence, unlike lithium. Hence, measured iron abundances should show less variation than lithium abundances in solar-type stars.
The contrast between lithium and iron appears in lower mass stars, where lithium depletion arises primarily from burning rather than downward mixing.
Since iron is not burned, there is no iron depletion in this mass range, even though lithium is depleted by many orders of magnitude. The iron enhancement is very small in low-mass stars, due to the larger masses of their surface convective zones.

\subsection{Engulfment Signature Timescales} \label{sec:timescale}

It is useful to quantify how long the engulfment signature is observable for comparison to spectroscopic observations of stars undergoing engulfment. We take this length of time to be the shortest of two candidate timescales.
The first timescale is how long it takes for the lithium enhancement to fall to less than 0.05 dex above an identical non-engulfment model. This is the level of natural chemical dispersion observed in stellar clusters, and thus the smallest achievable observational uncertainty to distinguish a lithium enriched or depleted star relative to a bound stellar companion \citep{de_silva2009}. The second timescale is when the lithium abundance of the engulfment model falls below the observable lithium abundance upper limit. Once this occurs, the lithium abundance is too low to directly observe, so abundance enhancements cannot be detected. The enrichment time scales for stars of different masses are listed in Table \ref{tab:table3}.

For low-mass stars ($\leq$0.7 $M_{\odot}$), the time span over which lithium enhancement can be detected is short and is strongly dependent on the star's mass, but not the engulfed mass. This is because the timescale is controlled by lithium burning near the base of the convective zone, which can deplete the surface lithium to below the observable limit.
Lower mass stars have deep convective zones, so this depletion occurs faster. The observable time scale increases from $\sim$10 Myr for $0.5 \, M_\odot$ stars to $\sim$1 Gyr for $0.7 \, M_\odot$ stars, and up to a few Gyr for $0.9 \, M_\odot$ stars. 

For stars with $M \! \gtrsim \! 0.9 \, M_\odot$, lithium is not rapidly destroyed by burning, but the size of the initial engulfment signature is small because of lithium already present on the star's surface prior to engulfment.  The initial burst of thermohaline mixing just after engulfment is enough to deplete this signature below 0.05 dex on the order of $10^7$ years or less. The models with less accreted mass have the shortest enrichment timescales because they have lower levels of initial lithium enrichment. Although the enrichment signature vanishes very quickly, we may observe a lithium depletion signature if the engulfment model becomes significantly lithium depleted relative to the non-engulfment model. This occurs in the 1.3 and 1.4 solar mass stars. In this mass range, exoplanet engulfment may be signaled by a lithium depletion signature, rather than lithium enhancement.

\begin{deluxetable*}{ccccccc}
\setlength{\tabcolsep}{1em}
\tablewidth{0.90\textwidth}
\tabletypesize{\footnotesize}
\tablewidth{0pt}
\tablecaption{Observable time scale of $^7$Li Enrichment Engulfment Signature \label{tab:table3}}
\tablecolumns{7}
\tablehead{
\colhead{Stellar Mass} &
\multicolumn{2}{c}{1 $M_{\oplus}$ Engulfment} &
\multicolumn{2}{c}{10 $M_{\oplus}$ Engulfment} &
\multicolumn{2}{c}{100 $M_{\oplus}$ Engulfment} \\ 
& \colhead{0.05 dex} & \colhead{Li Upper Limit}
& \colhead{0.05 dex} & \colhead{Li Upper Limit}
& \colhead{0.05 dex} & \colhead{Li Upper Limit}
}
\vspace{-20mm}
\startdata
0.5 $M_{\odot}$ &\textbf{9.1 Myr}&\textbf{9.1 Myr}&18 Myr&\textbf{8.8 Myr}&19 Myr&\textbf{8.5 Myr} \\
0.6 $M_{\odot}$ &1.1 Gyr&\textbf{120 Myr}&1.2 Gyr&\textbf{200 Myr}&910 Myr&\textbf{170 Myr}\\
0.7 $M_{\odot}$ & 4.7 Gyr&\textbf{520 Myr}&5.5 Gyr&\textbf{870 Myr}&4.7 Gyr&\textbf{640 Myr}\\
0.8 $M_{\odot}$ &8.1 Gyr&\textbf{740 Myr}&8.9 Gyr&\textbf{1.2 Gyr}&7.3 Gyr&\textbf{760 Myr}\\
0.9 $M_{\odot}$ &Never&\textbf{2.0 Gyr}&Never&\textbf{2.5 Gyr}&   Never&\textbf{1.6 Gyr}\\
1.0 $M_{\odot}$ &\textbf{Immediately}&10 Gyr&\textbf{54 Myr}&9.0 Gyr&   \textbf{110 Myr}&4.0 Gyr\\
1.1 $M_{\odot}$ &\textbf{Immediately}&Never&\textbf{38 Myr}&Never& \textbf{76 Myr}&Never\\
1.2 $M_{\odot}$ &   \textbf{0.6 Myr}&Never&   \textbf{340 Myr}&Never&   \textbf{48 Myr}&Never\\
1.3 $M_{\odot}$ &   \textbf{2.2 Myr}&2.5 Gyr&\textbf{240 Myr}&2.5 Gyr& \textbf{36 Myr}&2.1 Gyr\\
1.4 $M_{\odot}$ &   \textbf{1.7 Myr}&63 Myr&   \textbf{18 Myr}&87 Myr&   N/A*&N/A*\\
       \vspace{-2.5mm}
       
\tablecomments{Time span after planetary engulfment over which lithium enhancement can still be detected, when engulfment occurs at the start of the main sequence. The 0.05 dex timescale indicates when the lithium enhancement drops below 0.05 dex above a non-engulfment model. The Li Upper Limit timescale indicates when the lithium abundance of the engulfment model falls below the detection threshold. For a given combination, the shortest timescale is bolded. The 100 $M_{\oplus}$ accretion in the 1.4 $M_{\odot}$ model was unable to run to completion due to the large amount of accreted mass in the thin convective zone, causing problems with the model.}
\end{deluxetable*}



    
    

\section{Discussion} \label{sec:discussion}
In the very low-mass range for stars less than 0.7 $M_{\odot}$, we find that engulfment produces a detectable lithium enrichment signature that persists on a timescale much shorter than the stellar lifetime. This ranges from $\sim$1 Myr for 0.5 $M_{\odot}$ stars, to $\sim$1 Gyr for 0.7 $M_{\odot}$ stars. If we expect planetary engulfment to mostly occur near the beginning of a stellar lifetime, then it is unlikely that observed lithium enrichment in these stars is caused by planetary engulfment. Planetary engulfment would only be a likely explanation if the star is very young, or if there is ongoing steady accretion of debris disk material.

In the mass range from 0.8 to 0.9 $M_{\odot}$, the engulfment enrichment signatures can last for 10 Gyr or more, comparable to the age of a typical field star. However, the lithium abundance of the engulfment model drops below the level where we can only establish upper limits to lithium abundance via observations by about 1 Gyr. So, observed lithium enrichment in stars at the low end of this mass range could be explained by a relatively recent planetary engulfment event. It also means that we expect a large spread in observed Li abundances for stars in this mass range, depending on whether they have accreted planetary material or not. 
For stars between 1.0 and 1.2 $M_{\odot}$, the enrichment signature is very small, often below our cutoff of 0.05 dex. Thus, these signatures would be difficult to detect.

In the higher stellar mass range of 1.3 and 1.4 $M_{\odot}$ stars, our models exhibit significant lithium depletion for most of the MS lifetime, rather than lithium enhancement.  In this stellar mass range, the engulfment models may have surface lithium depletions by a factor of 10-100 relative to the non-engulfment model.  Over the range of planet masses we investigate (1-100 $M_{\oplus}$), the magnitude of this depletion does not strongly depend on the mass of the engulfed planet.
Planetary engulfment may manifest in observations as a separate population of stars that have significantly lower observed lithium abundances.  For example, in \citet{berger2018}, Figure 8 shows that within the $T_{\rm eff}$ range of $6000-6500$ K, most of the stellar population exhibits an $A({\rm Li})$ around 2.5, but there is a group of stars with abundances below an upper limit of $A({\rm Li}) \approx 1.5$.  

These predicted lithium enrichment patterns are similar to what is seen in large samples of stars with lithium abundance measurements such as \cite{aguileragomez2018,berger2018}. For most stars with $M \lesssim 1\, M_\odot$, the lithium abundance is too small to be measured. However, a small but significant fraction of these stars have higher lithium abundances by 1-2 orders of magnitude ($A({\rm Li})$ values of roughly 0.5-2.5), with lower mass stars exhibiting smaller abundances on average. This is very similar to the basic predictions of Figure \ref{fig:figure5} at $\sim$Gyr ages, so we posit that those lithium-enhanced stars engulfed a planet at some point in their past. Roughly $1.2 \, M_\odot$ stars typically have larger observed lithium abundances of $A(\rm{Li}) \sim 2.5$, similar to both our engulfment and no-engulfment models, so the two possibilities are difficult to distinguish. A small fraction of hot stars with $T>6000$K have small lithium abundances of $A(\rm{Li}) < 1.5$. These could arise from planetary engulfment as discussed above, or they could potentially arise from the descendants of lithium dip stars as proposed by \cite{aguileragomez2018}. 

\cite{spina2021} finds that $\sim$25\% of binary stars show chemical anomalies indicative of planetary engulfment. Most of the chemically anomalous pairs in that sample are in the $\sim \! 1.0-1.2 \, M_\odot$ mass range and show correlated lithium enhancements of $\Delta A(^7 {\rm Li}) \sim 0.5$ and $\Delta A({\rm Fe}) \sim 0.1$. While the iron enrichment levels roughly match our models, the lithium enhancements are larger than those shown for this mass range in Figures \ref{fig:figure5} and \ref{fig:figure7}. However, our models of engulfment events roughly 1 Gyr after ZAMS (Figure \ref{fig:figure8}) have about the right lithium enhancement. Hence, if those stars did engulf planets, the engulfment may have occurred long after the beginning of the main sequence.
Unfortunately, the precise masses of the engulfed planets are difficult to determine, because thermohaline mixing tends to equalize the surface enhancement, regardless of planetary mass (Figure \ref{fig:figure6}).

\cite{hawkins2020} find a similar fraction of chemically inhomogeneous wide binaries, with properties very similar to the chemically anomalous pairs of \cite{spina2021}. Hence, those inhomogeneous binaries may also result from planetary engulfment. \cite{oh2018} analyzed the Kronos \& Krios system, containing a wide pair of solar-like stars in which Kronos has a higher $[{\rm Fe/H}]$ by $\sim \! 0.2$ dex and a higher $[{\rm Li/H}]$ by $\sim \! 0.5$ dex. These are similar to our predicted values after the engulfment of a $\sim \! 10 M_\Earth$ planet by a $\approx \! 1.0 \, M_\odot$ star, perhaps requiring post-ZAMS engulfment as discussed above. We agree with the interpretation of \cite{oh2018} that the abundance difference of Kronos \& Krios arises from planetary engulfment. However, the mass of the engulfed planet depends sensitively on both the time since engulfment, and the convective envelope mass of Kronos.

\subsection{Comparison to Previous Studies}


 Our models include very similar physics to the investigation of \cite{theado2012}, and our conclusions are mostly the same, albeit with some important differences. 
That study found lithium depletion following engulfment in all masses they considered relative to pre-engulfment levels, with faster depletion in lower mass stars due to lithium burning.

However, \cite{theado2012} adopted the same composition for all their stellar models, with a pre-engulfment lithium abundance of $A({\rm Li}) \simeq 3.3$. This is slightly larger than observed in young solar-mass stars (Figure \ref{fig:figure1}), and orders of magnitude larger than expected for young low-mass stars. Our study evolved the star through the pre-MS with convection, so by the time our \texttt{MESA} stellar models reached the point of engulfment, their $A({\rm Li})$ values were significantly lower due to pre-MS lithium burning. $A({\rm Li})$ at engulfment ranged from $3$ for the 1.4 $M_{\odot}$ models to $-8$ for the 0.5 $M_{\odot}$ models.

For this reason, the accretion event hardly increased $A({\rm Li})$ in the \citet{theado2012} models, whereas our models show that it can increase by orders of magnitude, especially in low-mass stars. \cite{theado2012} then describe the lithium depletion that occurs after engulfment, but we emphasize that the surface lithium can still remain higher than its pre-engulfment level for many Gyr. Only in our most massive models ($M \gtrsim 1.2 \, M_\odot$) does the lithium abundance actually decrease relative to its pre-engulfment level, or relative to a model without planet engulfment.


The lithium depletion timescale of our models is similar to that derived by \cite{garaud2011} considering thermohaline mixing, which found surface reduction timescales of $\sim$10$^{7}$ years for $1.4 \, M_\odot$ stars. However, the interplay of gravitational settling and diffusion is also important in this mass regime due to the short diffusion timescales at the base of their thin convective envelopes. \citet{soaresfurtado2021} predicted lithium enrichment signatures following planetary engulfment, but they do not include thermohaline or diffusion processes. They derived survival times of 2.4 Gyr for a 1.0 solar mass star based on the lithium burning timescale, but we found that the survival time for the lithium enrichment can be much shorter, due primarily to thermohaline mixing.



\subsection{Limitations}  \label{sec:limitations}

There are several limitations in our study that can be used as areas for further investigation. One issue is how we handled mixing effects other than thermohaline mixing, overshoot mixing, and elemental diffusion. We represented these other mixing types with a constant \texttt{min\_D\_mix} coefficient and chose a value so that our model would fit observed surface lithium depletion in solar-like stars (Figure \ref{fig:figure1}). However this does not mean that our \texttt{min\_D\_mix} will match the behavior of these other mixing effects in all cases. An extra mixing that scales with the internal density (e.g., equation 1 of \citealt{dotter2017} or \citealt{deal2021}) may produce different results, especially for high-mass stars where the extra mixing just below the convective zone will be substantially larger.
Consequently, our predictions for $1.3 \, M_\odot$ and $1.4 \, M_\odot$ stars should be treated with caution, as higher envelope mixing (e.g., due to meridional circulation) could prevent depletion of refractory species due to gravitational settling. We encourage future studies to further examine the physics of these other mixing processes and their effects on lithium abundance following exoplanet engulfment.

Our model for pre-MS mixing and consequent lithium burning may not be accurate. We were forced to disable all mixing processes other than convection because they caused excessive lithium depletion in the pre-MS. Internal testing with our models confirmed that overshoot mixing is responsible. Other studies have noted difficulties with stellar models in predicting lithium depletion, particularly when considering convective overshoot mixing. \citet{fu2015} found that overshoot mixing sharply depletes lithium on the pre-MS, which they hypothesized is restored by late-stage residual mass accretion as the star approaches ZAMS. \citet{baraffe2017} proposed a new form for the convective overshoot diffusion coefficient within the region where overshoot mixing is active that varies as a function of stellar rotation rate, which we did not consider in our model.
 
Related to this issue, the pre-main sequence lithium depletion may be excessive in our models even after exclusion of the mixing processes besides convection.  Our 1 $M_{\odot}$ model has a surface lithium abundance slightly below the mean observed lithium abundance of solar-type stars at ZAMS \citep{somers2016}, demonstrated in Figure \ref{fig:figure1}. 
More robust models of mixing and lithium depletion on the pre-MS will lead to better models of engulfment signatures, which may be significantly influenced by the star's lithium abundance at the time of engulfment as evidenced by our comparison with \citet{theado2012}.

 In this study, we set the engulfment to occur at ZAMS, and only briefly investigated engulfment at later points during the lifetime of the star.
Protostars respond very differently than MS stars to planetary engulfment because their convective zones penetrate deeper into the star, where lithium is destroyed. Hence, pre-MS engulfments may produce a smaller engulfment signature than the MS engulfments we examined here. Conversely, engulfments occurring long after ZAMS may produce larger signatures, especially if gravitational settling has already created a stabilizing composition gradient that prevents thermohaline mixing. Future work should investigate this issue in more detail.


We also had to model planetary engulfment as a gradual accretion onto the stellar surface because that was how \texttt{MESA} could handle mass gain.  This differs from an actual planetary engulfment, which would have a duration of $\sim$years rather than the thousands of years our modeled accretion requires.
The planet could also plunge deep within the star before dissolution completes,  and slowly disintegrate over time \citep{jia2018}. In more massive stars, it is more likely that the engulfed planet can pass through the thin convective zone before breakup. These effects could lead to significantly different results in surface $^7$Li abundance evolution, e.g., lowering surface abundances, soon after engulfment. 
Consideration of these effects will lead to more accurate results from planetary engulfment models.

\section{Conclusion}

We used \texttt{MESA} to model planetary engulfment by stars at the start of the main sequence, and we then monitored the resulting evolution of surface $^7$Li abundance throughout the MS. Our models included thermohaline mixing, atomic diffusion, and convective overshoot. We found that the resulting $^7$Li engulfment signature is highly dependent on the stellar mass, and a strong function of time after the engulfment event. Our results are summarized as follows:

\begin{itemize}
\item In low mass stars below $\approx \! 0.7 \, M_{\odot}$, the engulfment signatures are short-lived. The convective zone reaches down into the lithium burning region where accreted lithium is quickly destroyed, and the surface lithium abundance is correspondingly reduced.

\item In solar-like stars between 0.7 and 0.9 $M_{\odot}$, the lithium enrichment signature is large, on the order of 0.5 dex or more, and can persist for $\gtrsim$1 Gyr.
These engulfment signatures are also insensitive to the amount of accreted mass, because an increase in engulfed mass is counteracted by the greater strength of thermohaline mixing below the convective zone.

\item At masses just above solar (1.0$-$1.2 $M_{\odot}$), the engulfment signatures are much smaller. In this mass regime, the initial lithium abundance without engulfment is already elevated relative to lower mass stars, so the engulfment does not increase the lithium abundance quite as much. Due to the combined action of thermohaline mixing and diffusion, increasing the mass of the engulfed planet does not produce a stronger and longer-lasting enrichment signature..

\item  At 1.3 and 1.4 $M_{\odot}$, the surface $^7$Li of stars undergoing engulfment becomes depleted relative to non-engulfment models. This occurs because the combination of thermohaline mixing and gravitational settling mix lithium below the convective zone. Higher mass stars have a thinner convective zone, thus the base of the convective zone is much closer to the stellar surface, and diffusion is much more effective. Thermohaline mixing allows lithium to be mixed deeper into the stellar interior, so stars that have undergone engulfment actually have less lithium at their surfaces compared to normal stars.

\item Surface lithium abundances are more enhanced for engulfment events occurring long after ZAMS, because thermohaline mixing is suppressed by stabilizing composition gradients that have formed via helium sedimentation.

\item Our predictions are broadly compatible with several abundance features apparent in large samples of stars. First, planetary engulfment can account for the large lithium enhancements that are observed in a fraction of cool ($T_{\rm eff} \lesssim 6000 \, {\rm K}$) stars. Second, planetary engulfment could potentially account for some hot stars with low lithium abundances. Third, measurements of correlated lithium/iron abundance differences in wide binary stars are at the right level to be explained by planetary engulfment.

\end{itemize}

\section*{Acknowledgments}
A.B. acknowledges funding from the National Science Foundation Graduate Research Fellowship under Grant No. DGE1745301. J.F. is thankful for support through an Innovator Grant from The Rose Hills Foundation, and the Sloan Foundation through grant FG-2018-10515. We thank Andrew Howard and Fei Dai for valuable input. 

\section*{Data Availability}
The data underlying this article will be uploaded to Zenodo.org  at \url{https://zenodo.org/communities/mesa} upon acceptance for publication.

\newpage
\bibliography{mybib}



\end{document}